\documentclass[aps,prl,paper,reprint,amsmath,amssymb,superscriptaddress,floatfix,showpacs,showkeys,subeqn,footinbib]{revtex4-1} 
\usepackage{amsmath, amsfonts, graphics, graphicx, hyperref,url,array, multirow, subfigure, rotating}
\usepackage{todonotes}
\usepackage{color}

\newcommand{\beq}{\begin{equation}}
\newcommand{\eeq}{\end{equation}}

\begin{document}

\title{Comment on "Explicit Analytical Solution for Random Close Packing in d=2 and d=3", Physical Review Letters {\bf 128}, 028002 (2022)}
\author{Raphael Blumenfeld}
\affiliation{Gonville \& Caius College, University of Cambridge, Trinity St., Cambridge CB2 1TA, UK}
\affiliation{ESE, Imperial College London, London SW7 2AZ, UK}

\date{\today}

\begin{abstract}
The method, proposed in \cite{Za22} to derive the densest packing fraction of random disc and sphere packings, is shown to yield in two dimensions too high a value that (i) violates the very assumption underlying the method and (ii) corresponds to a high degree of structural order. The claim that the obtained value is supported by a specific simulation is shown to be unfounded. One source of the error is pointed out.

\end{abstract}

\maketitle

%\section{\label{sec:Intro}I. Introduction}

A method is presented in~\cite{Za22} to evaluate the random close packing (RCP) fraction, $\phi_{RCP}$, of discs ($d=2$) and spheres ($d=3$). 
Two of the method's assumptions are: ``...$z=2d$ always applies at RCP..." and ``$z=2d$ is the only well defined criterion to define RCP", where $z$ is the mean coordination number per particle. I show that the method yields an impossibly high $\phi_{RCP}$ value at $d=2$, which  violates its very assumptions and corresponds to polycrystalline, rather than disordered, structure. The source of the problem is pointed out. \\

Following the exact analysis of ~\cite{Bl21}, consider the graph formed by connecting the centers of discs in contact in an $N(\to\infty)$-disc planar packing. It consists of $Nz/2$ edges and $N_c=(z-2)N/2$ cells~\cite{Euler}, which are the smallest polygons the edges make. Let a cell order be the number of discs in contact surrounding it and $Q_k$ the fraction of cells of order $k$ out of $N_c$~\cite{Waetal20}. The mean cell order, $\bar{k}$, and the mean coordination number are related~\cite{Bl21} by
\beq
z = \frac{2\bar{k}}{\bar{k}- 2} 
\label{ER}
\eeq
and the packing fraction can be expressed as~\cite{Bl21}
\beq
\phi = \frac{\pi (\bar{k}-2)}{8\sum_{k=3}^{K}Q_k\bar{S}_k} \ ,
\label{Phi}
\eeq
with $\bar{S}_k$ the mean area of cells of order $k$ and $K$ the highest cell order. 
The densest possible packings is crystalline, comprising only cells of order $3$ (equilateral triangles, $\bar{S}_3=\sqrt{3}/4$). To disrupt order, at least cells of order $4$ (rhombi, $\bar{S}_4=3/\pi$~\cite{Bl21}) must be included. Consider then packings comprising only triangles and rhombi (including higher order cells exacerbates the problem, as explained below). Using $\phi_{RCP}=0.88644$ in (\ref{Phi}), with $K=4$, yields $Q_3=0.873349$, a mean cell order $\bar{k}=3.126651$ and, using (\ref{ER}), $z = 5.550346$. This is significantly higher than $z = 2d = 4$, violating the very assumption on which the method in~\cite{Za22} is based.  \\

Additionally, consider the probability that a triangle is fully surrounded by triangles. It is the probability to find a triangle, $Q_3$, times the probability that it has exactly three triangular neighbors, $Q_3^3$, i.e., $Q_3^4=0.873349^4=0.581770$. So, more than $58\%$ of the triangles are {\it inside} crystalline clusters! 
The probability that a $3$-cell is surrounded by two triangles is $0.253101$, altogether more than $83\%$ triangles have either two or three triangular neighbors. Consequently, the occurrence probability of crystalline clusters decreases slowly with size, resulting in a polycrystalline packing. 
A good example of how a random packing even at $\phi=0.8681<0.88644$ unavoidably results in large polycrystals, was visualized in~\cite{LuSt90}. This example is annotated in~{SM} to emphasise the large extent of crystalinity and it makes it difficult to see how an even larger density could be arranged to appear disordered. To avoid this problem it is necessary to limit the probability of having two or three triangular neighbors to below $1/3$, which leads to a lower value of $\phi_{RCP}$~\cite{Bl21}.

Adding cells of orders $k>4$ to reduce the value of $z$ does not help. Since $\bar{S}_k$ increases with $k$, achieving $\phi_{RCP}=0.88644$ with $K>4$ necessitates increasing $Q_3$ to compensate. This increases further the area coverage of crystalline regions, exacerbating the ordering problem. \\

The origin of the error in~\cite{Za22} is the use of the Percus-Yevick and Carnahan-Starling approximations, which fail for jammed states~\cite{ChNi22}, while postulating that RCP occurs at $z=2d$. Moreover, in packings of frictionless discs/spheres, which are known to attain the highest density, $z=2d$ is the marginally stable state and, therefore, the loose, not the close, random packing.\\

In summary, the method proposed in \cite{Za22} yields too high a value of $\phi_{RCP}$ in $d=2$, leading to $z>5.5$, which violates the method's assumption $z=4$. It also corresponds to $58\%$ of the packing being crystalline, which means that a dense packing is polycrystalline, rather than disordered. Attempting to reduce $z$ by including higher-order cells only increases the degree of order. The error in \cite{Za22} stems from using approximations for the pair correlation function, which fail at jammed states, and the assumption $z=2d$, which is known to correspond to marginal rigidity and loose random packings. \\

{\bf Addendum:}\\

In a reply to this comment~\cite{Za22a}, Zaccone spotted a typo in my eq. (\ref{ER}) (corrected in this version), on the basis of which he concluded that my results have been derived from the incorrect equation. However, all the values derived in the comment, in particular that $\phi_{RCP}=0.88648$ results in $z>5.5$ and a crystalline coverage of more than $58\%$, were obtained with the {\it correct} equation and, hence, remain valid. Therefore, the above conclusions still stand: the method in \cite{Za22} is self-contradicting and leads to too high value of $\phi_{RCP}$, as calculated.\\

The reply~\cite{Za22a} contains several inaccuracies and misconceptions: that the approach, leading to the above results, is "heuristic", "contains several assumptions and approximations", and is based on a "peculiar protocol". 
(i) As detailed in \cite{Bl21}, the calculations are exact and no approximation is involved. (ii) The approach is based on no special assumption except that the packing and its statistics are spatially homogeneous. This assumption is simply Occam's razor -- to assume otherwise would be certainly more questionable. (iii) As explained clearly in~\cite{Bl21}, the conclusions are based on analysis of the cell order distribution and are valid for {\it all possible protocols} because any protocol must give rise to such a distribution. Thus, the conclusions in this comment are not protocol-specific, peculiar or otherwise. 
(iv) Zaccone combines the latter misconception with the argument: "it is well known that RCP in not a well defined concept precisely because, within a relatively broad range of packing fractions, different algorithms or protocols can generate packings with varying degree of structural order", to suggest that perhaps the result
$\phi_{RCP}=0.88648$ could be possible under some packing protocols. This argument also fails because the conclusions that the method in~\cite{Za22} leads to $z>5.5$ and to a crystalline coverage of more than $58\%$ are valid for {\it all possible protocols}.\\

Finally, the claim that the value $\phi_{RCP}=0.88644$ is supported by observations of $\phi=0.89$ in the simulations in~\cite{Meetal10} is misleading at best. Meyer {et al.}~\cite{Meetal10} obtained this value from a model, not a simulation algorithm. Moreover, Meyer {et al.} state explicitly that the value of $\phi_{RCP}=0.89$ must be wrong because it achieved in their model by ignoring particle overlaps. When they take the overlaps into consideration, they obtain $\phi_{RCP}\approx0.806$.

\end{document}